\documentclass[aps,twocolumn,showpacs,superscriptaddress,groupedaddress,nofootinbib]{revtex4} 
\usepackage[pdf]{pstricks}
\usepackage{graphicx}
\usepackage{graphicx}
\usepackage[sort&compress]{natbib}
\usepackage{subfigure}
\usepackage{amsmath}
\usepackage{amssymb}
\usepackage{amsfonts}
\usepackage{rotating}
\usepackage{cancel}
\usepackage{lipsum}
\usepackage{mathtools}
\usepackage{color}
\usepackage{bbm}
\usepackage{dsfont}
\usepackage{bbold}
\usepackage{multirow}
\usepackage{ulem}
\newcommand{\eps}{\epsilon}

\begin{document}

\title{The $T_{c\bar{s}}(2900)$ as a threshold effect from the interaction of the $D^*K^*$, $D^*_s\rho$ channels}

\author{R. Molina}
\author{E. Oset}
\affiliation{Departamento de F\'{\i}sica Te\'orica and IFIC,
Centro Mixto Universidad de Valencia-CSIC,
Institutos de Investigaci\'on de Paterna, Aptdo. 22085, 46071 Valencia, Spain}

\begin{abstract}  
We look at the mass distribution of the $D_s^+ \pi^-$ in the $B^0 \to \bar{D}^0 D_s^+ \pi^-$ decay, where a peak has been observed in the region of the $D^*_s \rho$, $D^* K^*$ thresholds. By creating these two channels  together with a $\bar{D}^0$ in $B^0$ decay and letting them interact as coupled channels, we obtain a structure around their thresholds, short of producing a bound state, which leads to a peak in the $D_s^+ \pi^-$ mass distribution in the $B^0 \to \bar{D}^0 D_s^+ \pi^-$ decay. We conclude that the interaction between the $D^*K^*$ and $D^*_s\rho$ is essential to produce the cusp structure that we associate to the recently seen $T_{c\bar{s}}(2900)$, and that its experimental width is mainly due to the decay width of the $\rho$ meson. The peak obtained together with a smooth background reproduces fairly well the experimental mass distribution observed in the $B_0 \to \bar{D}^0 D_s^+ \pi^-$ decay. 
\end{abstract}
\maketitle
\section{Introduction}
After the $X_0(2866)$, now called $T_{cs}(2900)$, by the LHCb Collaboration \cite{exp1,exp2}, in the $\bar{D}K$ spectrum of $B^+\to D^+D^-K^+$, many works have followed to explain this resonance from a compact tetraquark, sum rule derivations or molecular structure interpretations, among other (see references in \cite{daipap}). The molecular picture as a $\bar{D}^*K^*$ state studied from different perspectives, has obtained a broad support \cite{chen,chen2,Albuquerque,Liu,Huang,Hu,Xiao,Kong,Wang}. It is worth mentioning that such bound state was already predicted in \cite{branz} with properties very close to those observed experimentally. The $X_0(2900)$ as found in the work of \cite{branz} has $I=0$, and  $J^P=0^+$, being the latter in agreement with the quantum numbers associated to it in \cite{exp1,exp2}.

Interestingly, the $D^*K^*$ system was also investigated in \cite{branz} and three states were found corresponding to $I=0;J^P=0^+,1^+$ and $2^+$. The $2^+$ state was identified with the $D^*_{s2}(2573)$ state, and served to set the scale for the regularization of the loops, allowing predictions in the other sectors. There, the $I=1$ interaction of the $D^*K^*$ and $D^*_s\rho$ channels was also studied and, in section III-E, for $C(charm)=1;S(strangeness)=1$ and $I=1$, it was stated: ``For $J=0$ and $J=1$ we only observe a cusp in the $D^*_s\rho$ threshold''. This corresponds to a barely missed bound state, or virtual state. 

The recent finding by the LHCb Collaboration of a state observed in the $D^+_s\pi^-$, $D^+_s\pi^+$ mass distributions in the $B^0\to \bar{D}^0D^+_s\pi^-$ and $B^+\to D^-D_s^+\pi^+$ decays, respectively, at $2900$~MeV \cite{lhcbseminar}, gives us an incentive to reopen the issue and look at it from our prespective. Indeed, the state branded as $T_{c\bar{s}}(2900)$ with $J^P=0^+$, as seen in $D^+_s\pi^-$ and $D^+_s\pi^+$,  exhibits an $I=1$ character and it has also been associated with $J^P=0^+$. On the other hand, $2900$~MeV is just the threshold of the $D^*K^*$ channel. Thus, one is finding a $I=1$ $J^P=0^+$ state in the threshold of $D^*K^*$ (the $D^*_s\rho$ is only $14$ MeV below neglecting the $\rho$ width), which could correspond to the cusp found in \cite{branz}.

In the present work we look again at the interaction of $D^*K^*$ and $D^*_s\rho$ channels, taking into account the $K^*$ and $\rho$ widths and also the decay of the states found into the $D_s\pi$ channel where it has been observed, and compare our results with the experimental findings. 

We find a peak in the $D_s\pi$  distribution at the right place and a width in agreement with experiment, being the shape of the mass distribution also in good agreement with the experimental observation.
\section{Formalism}
\begin{figure}
 \begin{center}
  \begin{tabular}{c}
  \includegraphics[scale=0.5]{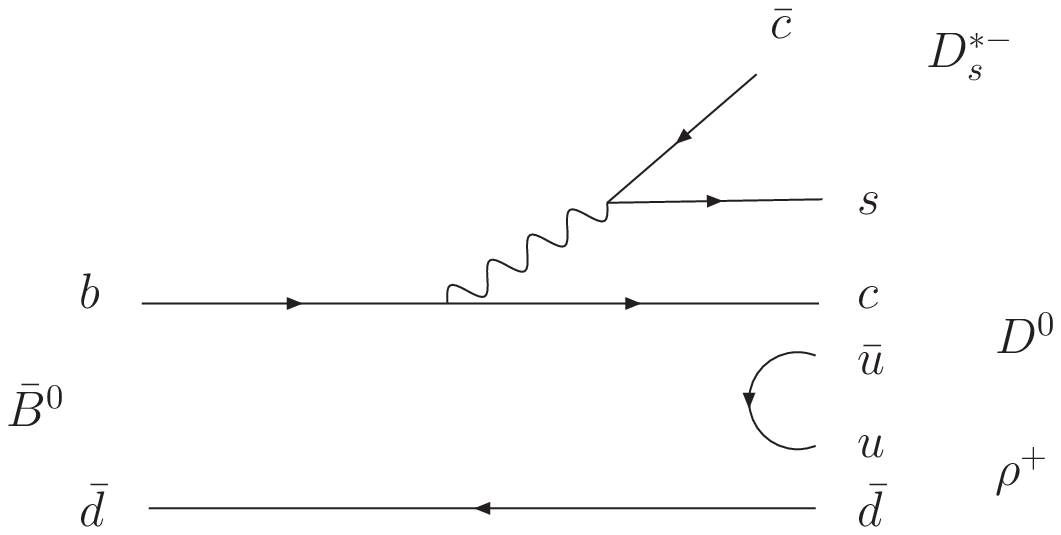}\\
  \includegraphics[scale=0.5]{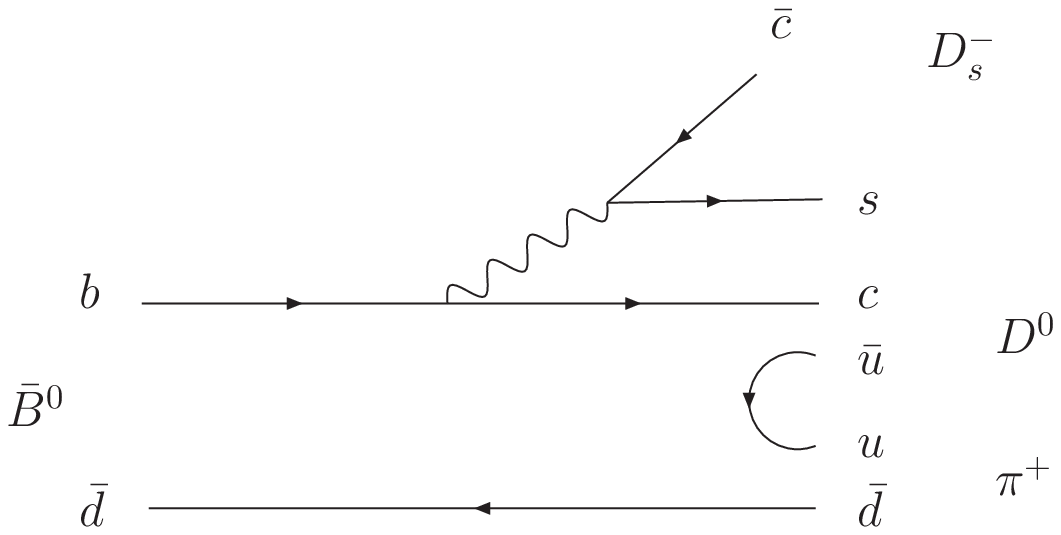}
  \end{tabular}
  \end{center}
  \caption{Top: $\bar{B}^0$ decay to $D^{*-}_sc\bar{d}$ with hadronization of the $c\bar{d}$ pair to produce $D^{*-}_sD^0\rho^+$. Bottom: $\bar{B}^0$ decay into $D_s^-D^0\pi^+$ (contribution to the background).}
  \label{fig:1}
 \end{figure}

 For $I=1$ in the sector with $charm$ (C) and $strangeness$ (S), $C=1;S=1$, we have two coupled channels, $D^*K^*$ and $D^*_s\rho$. It was shown in \cite{branz} that the system in $J=0$, as assumed in the experimental work, was barely short of binding but produced a cusp close to the energy of the two near by channels, $D^*_s\rho$ and $D^*K^*$. In Ref.~\cite{lhcbseminar} a peak is found in the $D_s\pi$ invariant mass in the $B^0\to\bar{D}^0D^+_s\pi^-$ and $B^+\to D^-D^+_s\pi^+$ decays. To visualize the process by means of which this decay can proceed, let us look at the $B$  weak decay at the quark level. In order to have a $b$ quark rather than a $\bar{b}$ quark, we look at the reaction $\bar{B}^0\to D^0D_s^{*-}\rho^+$. We produce this state with the external emission Cabibbo favored decay shown in Fig.~\ref{fig:1} (top). In Fig.~\ref{fig:1} (bottom) we depict the direct decay $\bar{B}^0\to D_s^-D^0\pi^+$ that we consider as background.
 
\begin{figure}
  \centering
  \includegraphics[scale=0.7]{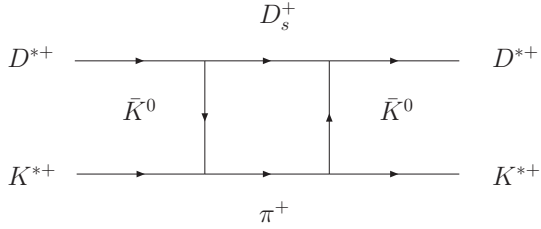}
  \caption{Box diagrams accounting for the $D^*K^*\to D^+_s\pi^+$ decay.}
  \label{fig:2}
 \end{figure}

 We produce $D^0D^{*-}_s\rho^+$ with $c\bar{d}$ hadronization with $\bar{u}u$, and $D^{*-}_s\rho^+$ forming an $I=1$ object. The direct production of the coupled channel $\bar{D}^*\bar{K}^*$ involves more complicated topological structures necessarily suppressed with respect to the $D^0D^*_s\rho^+$ production \cite{chau}. On the other hand, the $\pi D^-_s$ where the state is observed is not a coupled channel of the vector-vector ($VV$) channels that we have considered. It is a pseudoscalar-pseudoscalar ($PP$) decay channel which can be incorporated in the scheme via the box diagram of Fig.~\ref{fig:2}.

\begin{figure}
  \centering
  \includegraphics[scale=0.7]{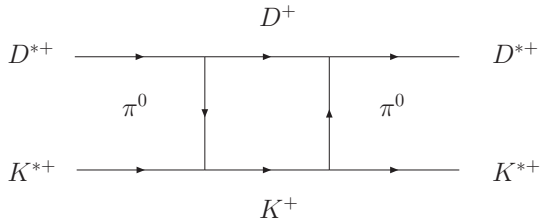}
  \caption{Box diagrams accounting for the $D^{*+}K^{*+}\to D^+K^+$ decay.}
  \label{fig:3}
 \end{figure}

\begin{figure}
  \centering
  \includegraphics[scale=0.55]{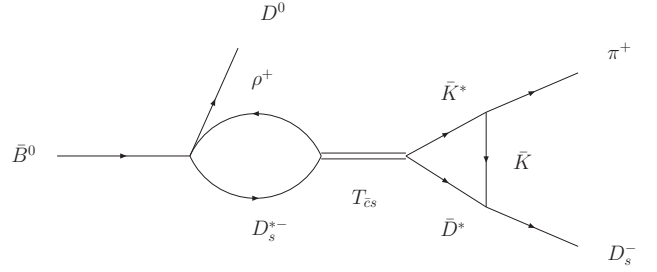}
  \caption{Mechanism by means of which the resonance is produced and decays into $\pi^+D_s^-$.}
  \label{fig:4}
 \end{figure}

 Still, we can have a more efficient decay channel $D^{*+}K^{*+}\to D^+K^+$, which is the one shown in Fig.~\ref{fig:3}.
 The smaller $\pi^0$ propagator in Fig.~\ref{fig:3} compared to the $K$ propagator in Fig.~\ref{fig:2} makes the source of imaginary part in the $VV$ potential more important for the mechanism of Fig.~\ref{fig:3}, which was evaluated in Ref. \cite{branz} also for the $D^*_s\rho$ channel. However, the state is observed in $D_s\pi$, hence, the mechanism by means of which the reaction proceeds is given in Fig.~\ref{fig:4}.
 
\begin{figure}
  \centering
  \includegraphics[scale=0.6]{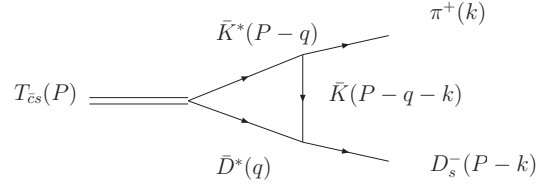}
  \caption{Triangle diagram accounting for the $R\to\pi \bar{D}_s$ decay of the $R$ resonance of $I=1$ generated with the $\rho\bar{D}_s$, $\bar{D}^*\bar{K}^*$ coupled channels.}
  \label{fig:5}
 \end{figure}

 The amplitude for the process of Fig.~\ref{fig:4} is given by,
 \begin{equation}
  t=aG_{\rho D^{*}_s}(M_\mathrm{inv})t_{\rho D^*_s,K^*D^*}(M_\mathrm{inv})\tilde{V}(\pi D_s, M_\mathrm{inv})\label{eq:t}
 \end{equation}
 where $a$ is a normalization constant that we do not evaluate, unnecessary to show the shape of the $\pi D_s$ mass distribution in the $\bar{B}^0$ decay, and $M_\mathrm{inv}$ is the invariant mass distribution of the $D_s\pi$ final state. The vertex function $\tilde{V}$ corresponding to the triangle loop of Fig.~\ref{fig:5} can be easily evaluated. Note that in principle we should also consider the $t_{\rho D_s^*\to \rho D_s^*}$ transition, but the triangle loop with $D_s^*\rho$ intermediate state, with a $\pi$ replacing the $K$, is zero because $D_s^*$ and $D_s$ have no overlap with the $u,d$ quarks of the pion.
 
 Since any normalization of the triangle diagram can be incorporated in the coefficient $a$ of Eq.~(\ref{eq:t}), we do not care about the values of the vertices but only about their structure,
 \begin{eqnarray}
  &&\bar{K}^*\to \pi\bar{K}:\, \vec{\eps}_{K^*}\cdot(2\vec{k}-\vec{P}+\vec{q})\nonumber\\
  &&\bar{D}^*\to \bar{D}_sK:\, \vec{\eps}_{D^*}\cdot(2\vec{P}-\vec{q}-2\vec{k})\nonumber\\
  &&R\to\bar{K}^*\bar{D}^*: \,\vec{\eps}_{\bar{K}^*}\cdot\vec{\eps}_{D^*}\ .\label{eq:ver}
 \end{eqnarray}
We have assumed the resonance to be in $J=0$, hence the $\vec{\eps}_{\bar{K}^*}\vec{\eps}_{\bar{D}^*}$ coupling, and we have also assumed that the vectors have small momenta with respect to their masses, which is true when $\bar{K}^*$, $\bar{D}^*$, are close to on-shell in the loops from where the largest contribution to the vertex comes in the integration. This allows us to neglect the $\eps^0$ component of the vectors. We take $\vec{P}=0$, in the $\pi\bar{D}_s$ rest frame and then the structure of the triangle diagram of Fig.~\ref{fig:5} is given by
\begin{eqnarray}
 &&\tilde{V}=-i\int \frac{d^4q}{(2\pi)^4}\eps_{\bar{K}^*}^l\eps_{\bar{D}^*}^l\eps_{\bar{K}^*}^i\eps_{\bar{D}^*}^j\frac{(2k+q)^i(2k+q)^j}{(P-q-k)^2-m^2_K+i\eps}\nonumber\\&&\times\frac{\theta(q_\mathrm{max}-q)}{(P-q)^2-m_{K^*}^2+i\eps}\frac{1}{q^2-m^2_{D^*}+i\eps}\ .\label{eq:vt}
\end{eqnarray}
The loop function $\tilde{V}$ is naturally regularized with a cutoff $q_\mathrm{max}$, the same one used to regularize the $D^*K^*$ and $D^*_s\rho$ loops when studying their interactions. This can be seen  since the coupled channel approach with a cutoff regularization is equivalent to using a separable potential $V\theta(q_\mathrm{max}-q)\theta(q_\mathrm{max}-q')$, which leads to a separable $t$ matrix, $t\theta(q_\mathrm{max}-q)\theta(q_\mathrm{max}-q')$ \cite{danijuan}, in this case, $t_{\rho D^*_s,K^*D^*}$ of Eq.~(\ref{eq:t}). The equivalent $q_\mathrm{max}$ used in \cite{branz} was $1100$~MeV. 

We split the propagators into the positive and negative energy parts as,
\begin{equation}
 \frac{1}{q^2-m^2+i\eps}=\frac{1}{2\omega(q)}\left(\frac{1}{q^0-\omega(q)+i\eps}-\frac{1}{q^0+\omega(q)-i\eps}\right)\ ,
\end{equation}
with $\omega(q)=\sqrt{\vec{q}\,^2+m^2}$, and keep only the positive energy part for the heavy mesons $\bar{D}^*$, $\bar{K}^*$, retaining the two terms for the kaon propagator. The $q^0$ integration is then easily done using Cauchy's residues and, after summing over the internal $K^*$, $D^*$ polarizatios, we find,

\begin{eqnarray}
 &&\tilde{V}=-\int\frac{d^3q}{(2\pi)^3}\frac{(2\vec{k}+\vec{q})^2}{8\omega_{K^*}(q)\omega_{D^*}(q)\omega_K(\vec{q}+\vec{k})}\nonumber\\&&\times\frac{1}{P^0-\omega_{D^*_s}(q)-\omega_{K^*}(q)+i\eps}\nonumber\\&&\left\{\frac{1}{P^0-k^0-\omega_{D^*}(q)-\omega_K(\vec{q}+\vec{k})+i\eps)}\right.\nonumber\\&&\left.+\frac{1}{k^0-\omega_{K^*}(q)-\omega_K(\vec{q}+\vec{k})+i\eps}\right\}\ ,
\end{eqnarray}
which shows the different cuts of the loop diagram when pairs of the internal particles of the loop are placed on-shell.

Then, we consider that the transition amplitude for $\bar{B}^0\to D^0D^-_s\pi^+$ is given by a constant background (considering the dominance of s-wave in the coupling of the bottom meson to the pseudoscalars), see Fig.~\ref{fig:1} (bottom), together with the scattering amplitude of the diagram in Fig.~\ref{fig:4}, which accounts for the interaction of the $VV$ coupled channels. It reads as
\begin{equation}
  t'=aG_{\rho D^*_s}(M_\mathrm{inv})t_{\rho D^*_s,K^*D^*}(M_\mathrm{inv})\tilde{V}(\pi D_s, M_\mathrm{inv})+b\label{eq:t1}
 \end{equation}

Therefore, the mass distribution of $\pi D_s^-$ in the $\bar{B}^0$ decay is given by,
\begin{equation}
 \frac{d\Gamma}{dM_{\mathrm{inv}}}=\frac{1}{(2\pi)^3}\frac{1}{4M^2_B}p_{D^0}\tilde{p}_\pi\vert t'\vert^2\ ,\label{eq:dg}
\end{equation}
where \begin{eqnarray}
  p_{D^0}=\frac{\lambda^{1/2}(M^2_B,m^2_{D^0},M^2_{\mathrm{inv}})}{2M_B};\quad    \tilde{p}=\frac{\lambda^{1/2}(M^2_\mathrm{inv},m^2_{D_s},m^2_\pi)}{2M_\mathrm{inv}}\ .  \nonumber
      \end{eqnarray}

\section{Results}
The different contributions to the potential for the case of $C=1;S=1;I=1$ and $J^P=0^+$ are given in Table XIV of \cite{branz}. We notice that, for the $D^*K^*\to D^*K^*$ tree-level amplitude, contrary to the case of the $T_{cs}(2900)$, where the interaction driven by $\rho$-exchange was three times bigger than for $\omega$-exchange, these two exchanges have similar strengths in this sector but also opposite sign, and therefore, the interaction is negligible in this transition element. Being this element also zero for $D^*_s\rho\to D^*_s\rho$ due to the OZI rule. Instead we get a relatively large transition potential for $D^*_s\rho\to D^*K^*$. The situation with two channels where the diagonal elements of the potential are null but there is an appreciable non-diagonal transition potential appears often in hadronic physics problems. The existence of this transition potential $V_{12}$ when $V_{11}, V_{22}$ are zero acts as a source of attraction in channel $1$. Indeed, it is shown in Sec.~6 of \cite{aceti} that one can eliminate channel $2$ and obtain the same amplitude $t_{11}$ using an effective potential in one channel, $V_{\mathrm{eff}}=V_{11}+V_{12}^2G_2$, and since $\mathrm{Re}G_2<0$ the new term acts as an attractive potential. Thanks to that, one can obtain the $\Omega(2012)$ state from the coupled channels $\pi\Sigma^*$, $\eta\Omega$, with null diagonal potentials \cite{kolo,sarkar,pavao,xie,pavon}, and a cusp like structure for the $Z_{cs}(3985)$ from the interaction of the $D_s^*\bar{D}^*$ and $J/\psi K^*$ channels \cite{ikenoraquel}. 

\begin{figure}
 \begin{center}
 \begin{tabular}{c}
  \includegraphics[scale=0.35]{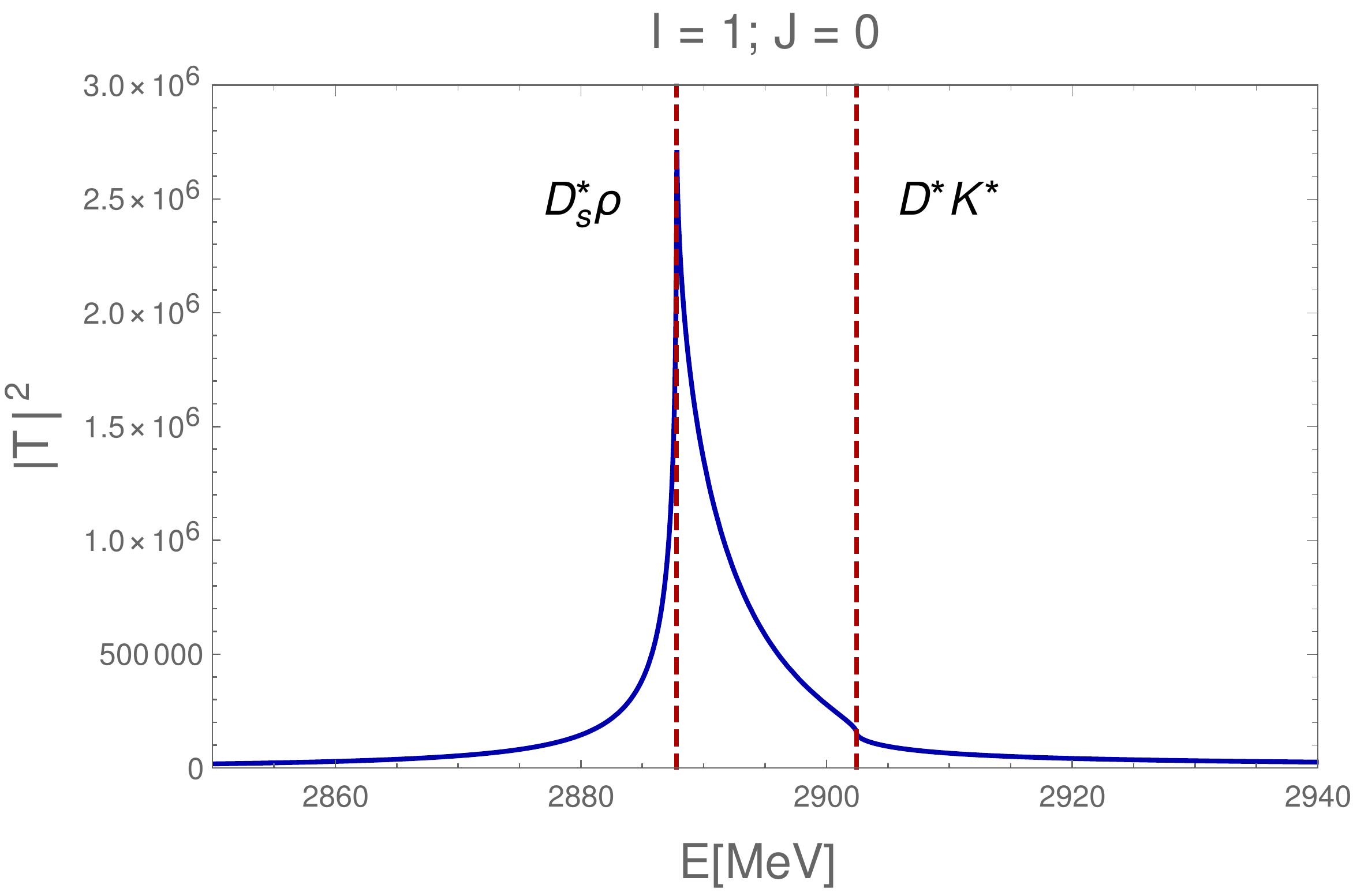}\\\includegraphics[scale=0.35]{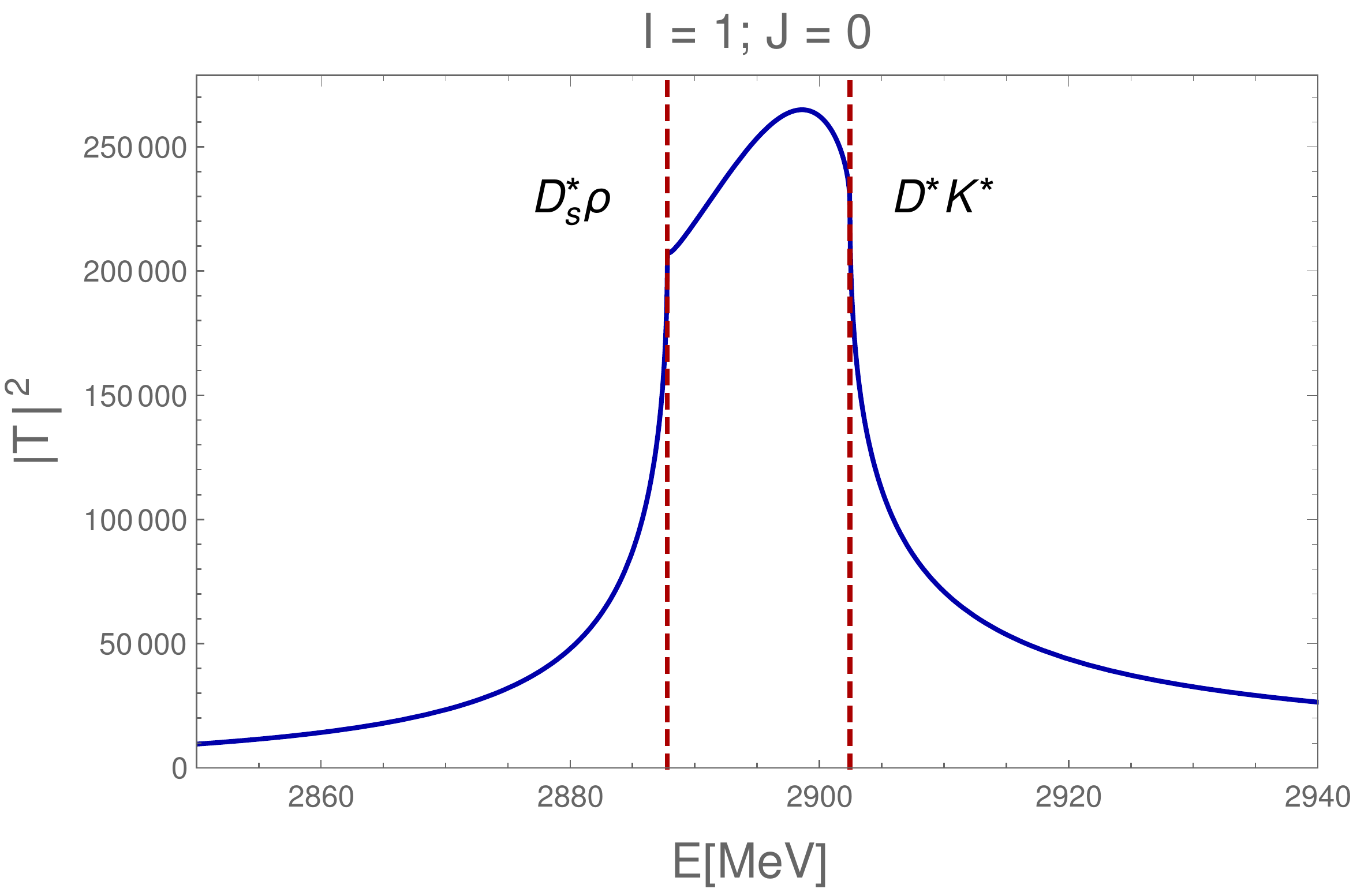}
 \end{tabular}
\end{center}
\caption{Results with the potential of Table~XIV of \cite{branz}, and including also the box diagram of Fig.~\ref{fig:3}, with the parameters used in Ref.~\cite{branz}, $\alpha=-1.6$, $\Lambda=1200$ (top), and with the new parameters fixed to obtain the $T_{cs}(2900)$ \cite{raquel}, $\alpha=-1.474$, and $\Lambda=1300$ MeV (bottom).}
\label{fig:cusp}
\end{figure}

For an illustration we show first the results with the same parameters used in \cite{branz}, $\alpha=-1.6$, $\Lambda=1200$~MeV in Fig.~\ref{fig:cusp} (top) (not shown in \cite{branz}), where the tree-level amplitudes of Table~XIV of \cite{branz} and the box diagram with intermediate $DK$ in the $D^*K^*$ channel, Fig.~\ref{fig:3}, are included for $I=1;J=0$. As discussed in \cite{branz}, a cusp is obtained in the $D^*_s\rho$ threshold. The fact that there is not a sharp cusp near the $D^*K^*$ threshold is related to the box diagram of Fig.~\ref{fig:3} which allows for the decay into $DK$. Since we have now the new information of the $T_{cs}(2900)$ mass and decay width, we can slightly adjust the parameters in order to reproduce them. This was done in \cite{raquel}, obtaining $\alpha=-1.474$ and $\Lambda=1300$~MeV. With this new set of parameters we plot $\vert T\vert^2$ for $C=1;S=1;I=1$ in Fig.~\ref{fig:cusp} (bottom). We still obtain a cusp but now the strength of the peak accummulates more around the $D^*K^*$ threshold. 

\begin{figure}
\begin{center}
 \includegraphics[scale=0.35]{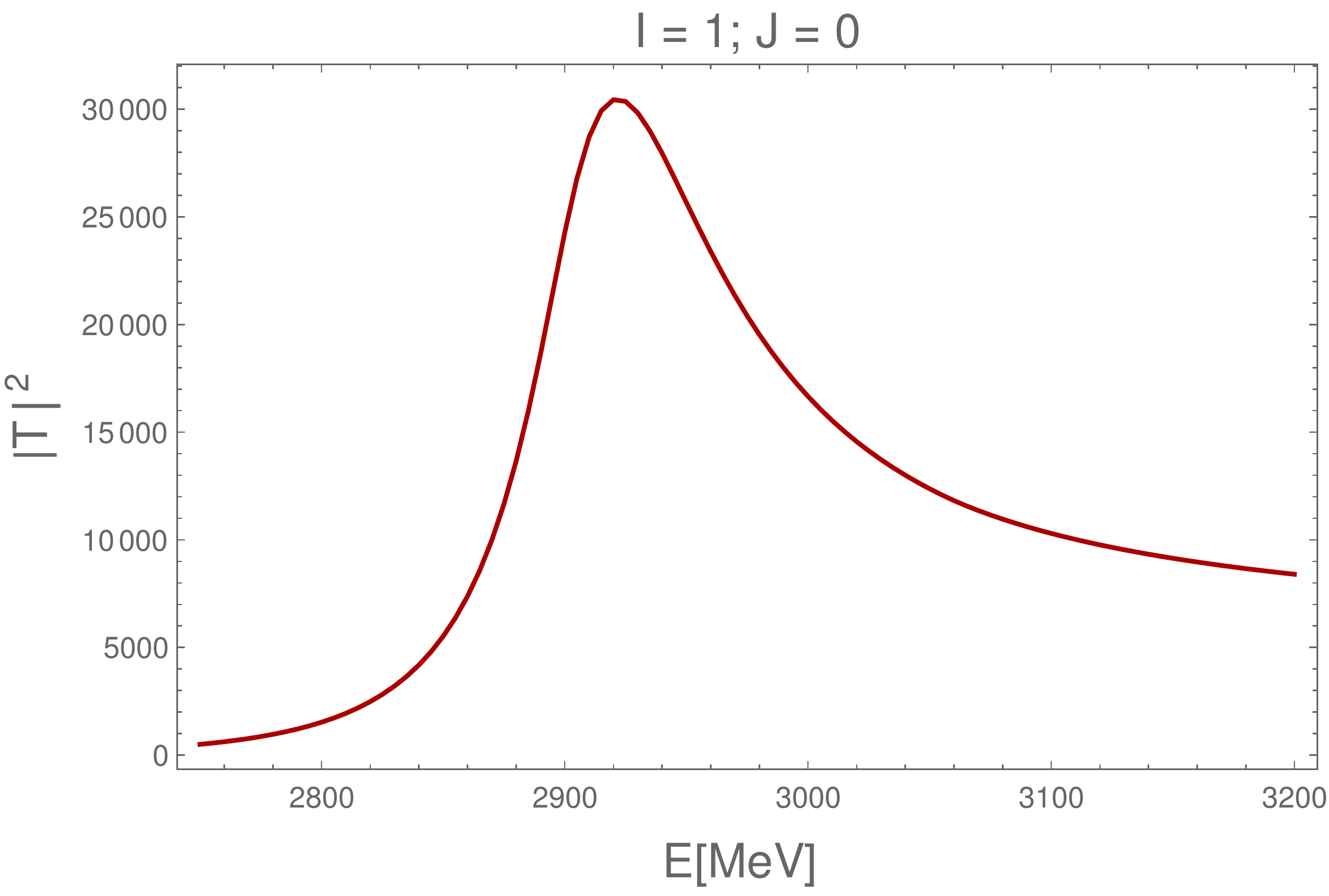}
 \end{center}
 \caption{$\vert T\vert^2$ for $C=1;S=1;I=1;J=0$ with $\alpha=-1.474$.}.
 \label{fig:t101}
\end{figure}

It is clear that even though the peak is already visible around the position seen in the experiment, the width obtained (around $16$~MeV in Fig.~\ref{fig:cusp} (bottom)) is much narrower than the observed one. Next, we consider the decay width of the $\rho$ and $K^*$ mesons by means of the convolution of the two meson loop function with an energy dependent width,
\begin{eqnarray}
\tilde{G}(s)&=& \frac{1}{N}\int^{M_\mathrm{max}^2}_{M_\mathrm{min}^2}d\tilde{m}^2_1(-\frac{1}{\pi}) {\cal I}m\frac{G(s,\tilde{m}^2_1,M_2^2)}{\tilde{m}^2_1-M^2_1+i\Gamma(\tilde{m})\tilde{m}_1} \ ,\nonumber
\label{Gconvolution}
\end{eqnarray}
with
\begin{equation}
N=\int^{M_\mathrm{max}^2}_{M_\mathrm{min}^2}d\tilde{m}^2_1(-\frac{1}{\pi}){\cal I}m\frac{1}{\tilde{m}^2_1-M^2_1+i\Gamma(\tilde{m})\tilde{m}_1}\ ,
\label{Norm}
\end{equation}
where $M_1$ is the nominal mass of the vector meson,  $M_{\mathrm{min}}=M_1-3.5\Gamma_0$, $M_{\mathrm{min}}=M_1+3.5\Gamma_0$, with $\Gamma_0$ the nominal mass of the $\rho$ and $K^*$ mesons, and
\begin{equation}
\tilde{\Gamma}(\tilde{m})=\Gamma_0\frac{q^3_\mathrm{off}}{q^3_\mathrm{on}}\Theta(\tilde{m}-m_1-m_2)
\end{equation}
with
\begin{equation}\label{eq:mom}
q_\mathrm{off}=\frac{\lambda^{1/2}(\tilde{m}^2,m_1^2,m_2^2)}{2\tilde{m}},\quad
q_\mathrm{on}=\frac{\lambda^{1/2}(M_1^2,m_1^2,m_2^2)}{2 M_1}\ ,
\end{equation}
where $m_1=m_2=m_\pi$ for the $\rho$, and $m_1=m_K,m_2=m_\pi$ for the $K^*$. The result when we take into account the decay widths of the vector mesons is plotted in Fig.~\ref{fig:t101}. Now the cusp obtained for $J=0$ has  softened because of the consideration of the decay widths of the vector mesons. The position of the cusp is similar, it shows up slightly above the $D^*K^*$ threshold and around $2920$ MeV, with a width coming basically from the decay of the $\rho$ into $\pi\pi$. We do not find any pole in the second Riemann sheet. All the results shown here have been evaluated using ``model B'' for the box diagram in \cite{branz} with $\Lambda=1300$ MeV as in \cite{raquel}. We notice that the results are practically the same for $\Lambda=1200$~MeV and $1300$~MeV. Most of the width comes in this case from the decay of the vector mesons instead. These results are summarized in Table~\ref{tab:pos}, where we also include for completeness what we obtain with the present input for $J=1$ and $J=2$ \cite{branz}.

\begin{figure}
 \centering
 \includegraphics[scale=0.35]{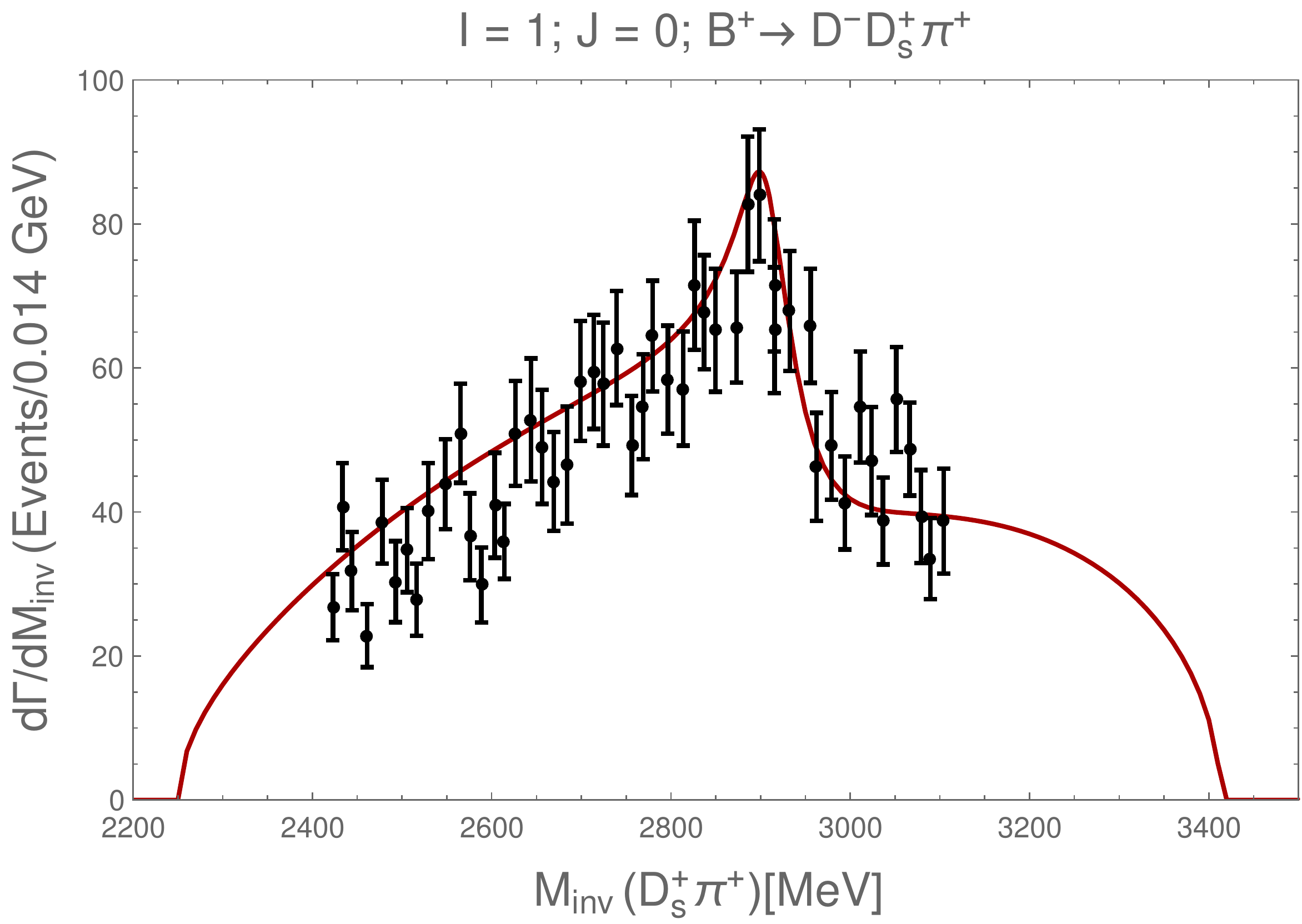}
 \caption{Invariant mass distribution for $D_s\pi$ from the decay $B\to \bar{D}D_s\pi$ compared to the experimental data from Ref.~\cite{lhcbseminar}.}
 \label{fig:compexp}
\end{figure}

\begin{table}
\renewcommand{\arraystretch}{1.6}
 \setlength{\tabcolsep}{0.2cm}
\begin{center}
\begin{tabular}{ccccc}
 $I[J^{P}]$&\multicolumn{2}{c}{$\sqrt{s_0}$}&$\Gamma_0$&\multicolumn{1}{c}{Experiment} \\
 \hline
 \hline
 $1[0^+]$&$2920$&(Cusp)&$130$&$m=2908\pm 11\pm 20$\\
 &&&&$\Gamma=136\pm 23\pm 11$ \\
 $1[1^+]$&$2923$&(Cusp)&$145$&-\\
 $1[2^+]$&\multicolumn{2}{c}{$2834$}&$19$&-\\
 \hline
\end{tabular}
 \end{center}
\caption{Position and width of the cusp/state obtained in comparison with the experiment.}
\label{tab:pos}
\end{table}
Finally, we show the result of the invariant mass distribution of the decay $\bar{B}^0\to D^-_s D^0\pi^+$, Eq.~(\ref{eq:dg}), in comparison with the LHCb experimental data \cite{lhcbseminar} in Fig.~\ref{fig:compexp}\footnote{We compare with the data of the $D^+_s\pi^+$ mass distribution in the $B^+\to D^-D^+_s\pi^+$ analogous decay of \cite{lhcbseminar}, where the peak is clearly seen.}. In Eq.~(\ref{eq:dg}), we adjusted the constants $a$ and $b$ to reproduce well the experimental data around the $T_{c\bar{s}}(2900)$ resonance, and we obtain $a=2.1\times 10^3$ and $b=-1.5\times 10^3$. As can be seen, our model describes well the experimental data. A peak is obtained around the threshold of the $D^*K^*$ channel and a sharp dip, caused by the interference between the triangle loop in Fig.~\ref{fig:4}, the cusp obtained in the scattering amplitude shown in Fig.~\ref{fig:t101}, and the background. Since these results where obtained fixing the subtraction constant to obtain the $T_{cs}(2900)$, this also supports the molecular picture of this state as $D^*\bar{K}^*$ of \cite{raquel}. Thus, our model strongly supports the $T_{c\bar{s}}(2900)$ as a cusp structure originated by the non-diagonal interaction $D^*K^*\to D^*_s\rho$, with a width mainly due to the decay of the $\rho$ meson into $\pi\pi$

\section{Conclusions}
We have studied the $B^0 \to \bar{D}^0 D^+_s \pi^-$ decay in the region of the $D^*_s \rho$, $D^* K^*$ masses, by considering explicitly the interaction of these two coupled channels within the framework of the local hidden gauge approach. A peak is observed experimentally in the $D_s^+ \pi^-$ mass distribution that we associate to the structure created by the production of the $D^{*+}_s \rho$ channel in the $B^0 \to \bar{D}^0 D^{*+}_s \rho^-$ decay followed by a transition $D^{*+}_s \rho^-$ to  $D^* K^*$ which decays finally to $D_s^+ \pi^-$.  The process involves the interaction of $D^{*+}_s \rho^-$, $D^* K^*$ coupled channels in isospin $I=1$, $J^P=0^+$, which is relatively weak but creates a threshold structure. Indeed, the diagonal interaction terms of this system are null, but the transition potential between the two channels acts as an attraction, short of binding, but which gives rise to a strong cusp. When the widths of the $\rho$ and $K^*$ are considered, this cusp gives rise to a peak structure in very good agreement with the experimental findings.
The peak can be considered as a virtual state created by the  $D^{*}_s \rho$, $D^* K^*$ interaction in coupled channels.

\section{Acknowledgments}
R. M. acknowledges support from the CIDEGENT program with Ref. CIDEGENT/2019/015 and from the spanish national grant PID2019-106080GB-C21.
This work is also partly supported by the Spanish Ministerio de Economia y Competitividad and European FEDER funds under Contracts No. FIS2017-84038-C2-1-P B and No.  FIS2017-84038-C2-2-P B. This project has received funding from the European Union’s Horizon 2020 research and innovation programme under grant agreement No. 824093 for the STRONG-2020 project.

\bibliography{bibliotc}

\end{document}